\documentclass[12pt]{article}
\usepackage{amsfonts}
\usepackage{fancyhdr}
\usepackage{comment}
\usepackage{natbib}
\usepackage[a4paper, top=2.5cm, bottom=2.5cm, left=2.2cm, right=2.2cm]{geometry}
\usepackage{times}
\usepackage{amsmath}
\usepackage{changepage}
\usepackage{colortbl}
\usepackage{amssymb}
\usepackage{graphicx}
\usepackage{tikzsymbols}
\newtheorem{theorem}{Theorem}

\newtheorem{definition}[theorem]{Definition}

\begin{document}
\bibliographystyle{apsr}

\title{Algorithmic Fairness and Statistical Discrimination}
\author{John W. Patty\thanks{Professor of Political Science and Quantitative Theory \& Methods, Emory University,  \textit{jpatty@emory.edu}.} \and
	Elizabeth Maggie Penn\thanks{Professor of Political Science and Quantitative Theory \& Methods, Emory University,  \textit{empenn@emory.edu}.}}
\date{\today}

\maketitle

\begin{abstract}
    Algorithmic fairness is a new interdisciplinary field of study focused on how to measure whether a process, or \textbf{algorithm}, may unintentionally produce unfair outcomes, as well as whether or how the potential unfairness of such processes can be mitigated. Statistical discrimination describes a set of informational issues that can induce rational (\textit{i.e.}, Bayesian) decision-making to lead to unfair outcomes even in the absence of discriminatory intent.  In this article, we provide overviews of these two related literatures and draw connections between them.  The comparison illustrates both the conflict between rationality and fairness and the importance of endogeneity (\textit{e.g.}, ``rational expectations'' and ``self-fulfilling prophecies'') in defining and pursuing fairness.  Taken in concert, we argue that the two traditions suggest a value for considering new fairness notions that explicitly account for how the individual characteristics an algorithm intends to measure may change in response to the algorithm.
    
\end{abstract}

\newpage

\section{Introduction}

Algorithms are increasingly important to many, if not most, facets of everyday life.  Algorithms determine not only how resources and opportunities---such as employment, housing, credit, and education---are distributed, but also dictate both the options available to choose from and the information available about those options.  Furthermore, aided by the availability of vast individual-level data and cheap computational resources, these increasingly ubiquitous algorithms are also increasingly opaque and complex.  This combination of pervasiveness and inscrutability has led to concerns about the social impacts of algorithms.\footnote{\cite{KleinbergMullainathan19} offer an insightful and powerful analysis of the potential costs of pursuing simplicity in these algorithms.  As with all of the works cited herein, this work has greatly facilitated our thinking through many issues in this new literature.}

\paragraph{Algorithmic Fairness.} Algorithmic fairness (\textit{AF}) is a new term describing the study of how to evaluate rule-based procedures for making decisions about diverse individuals.  At the heart of this study is the presumption that certain ways of discriminating between two or more individuals are undesirable (\textit{i.e.}, ``unfair''), whereas others are less suspect, or even desirable (\textit{i.e.}, ``permissible'').  This field has quickly emerged as an active, important, and multidisciplinary research agenda over the past 20 years.\footnote{Issues of AF are being raised and studied in numerous academic fields, including medicine, criminology, philosophy, law, economics, and computer science.}  Driving this emergence is the increased use of sophisticated algorithms in various settings that affect people on an everyday basis.  

The principles underlying measures of AF are typically statistical, with a focus on different notions of an algorithm's \textbf{accuracy}.  As we discuss in more detail below, an algorithm $A$ is typically defined to be more ``fair'' than another, $B$, if $A$ is more ``equally accurate'' with respect to people from different groups (\cite{Sharifi19}), frequently defined by one or more ``protected traits,'' such as race, gender, or sexual orientation.  In general, regardless of the fairness metric one chooses, AF seeks to design an algorithm such that its errors do not systematically disadvantage one group of people relative to another.

\paragraph{Statistical Discrimination.} The literature on statistical discrimination (\textit{SD}) is more established than that on AF.  Rather than measuring and classifying disparities in algorithmic performance across groups, this literature squarely aims to identify the root causes of discrimination, and to disentangle disparate outcomes due to discrimination (\emph{i.e.}, \emph{disparate treatment}) from those due to exogenous disparities across groups (\cite{Lang20}).  Theories of SD typically assume that a decision-maker, such as an employer, makes decisions based on observable traits of individuals (\textit{e.g.} gender or race) that may be correlated with unobservable but outcome-relevant traits, such as skill.\footnote{Theories of \textit{statistical} discrimination differ from those of \textit{taste-based} discrimination (\textit{e.g.}, \cite{Becker71}) in that the discrimination arises from valid statistical inference and not from any animus toward particular groups.  That said, the fact that a decision-maker may have no preferential bias toward or against any group does not mean that the outcomes stemming from a statistically discriminatory process are any less harmful than those stemming from taste-based discrimination.}

Our goal in this article is to introduce the reader to these two, methodologically distinct, approaches to the study of discrimination and to highlight the theoretical connections between the fields. The AF literature takes an exogenous collection of individuals and traits as a primitive and aims to design a sorting mechanism that balances predictive accuracy with fairness.  Theoretical work on SD takes a strategic decision-maker and strategic agents as primitives, and aims to understand why inequality can emerge in various decision-making environments. Clearly the approaches differ in the agency afforded to the individuals being evaluated, and in how decisions about individuals are conceptualized (via a mathematical formula versus the profit-maximizing actions of an employer, for example).  The most significant difference between the two approaches is in how they conceive of the outcome-relevant traits of individuals that the algorithm or decision-maker seeks to learn (again, such as the skill of an applicant). In the AF literature these individual-level traits are typically taken as exogenous. In theories of SD they are often the product of investment by the people being evaluated, and therefore endogenous to the actions of the decision-maker.

\subsection{Distinguishing Between Algorithmic Fairness \& Statistical Discrimination}  While the principal goal of this article is to clarify the connections between AF and SD, it is important to note up front that the two are neither equivalent nor ``logically nested'': there are questions in SD that have nothing to do with AF, and vice-versa.   To make this clear, we first walk through a contextualized hypothetical ``employment'' example and then turn to a recent, real world example of a policy question related to AF, but not SD  in a non-employment context.

Suppose that applicants for a job are from two different groups, ``male'' and ``female.''  Every applicant is either qualified or not, but this is not directly observable.  Rather, each applicant has taken a \textbf{test}, and the result of this test for applicant is positively correlated with whether he or she is qualified.  To make things concrete, suppose that the test is scored on a 0-100 point scale.  The employer can observe both the applicant's test score and his or her group membership, and suppose that the employer hires any applicant from group $g \in \{\text{male},\text{female}\}$ if and only if his or her test score is greater than or equal to the employer's \textbf{threshold} for group $g$, denoted by $t(g) \in \{0,\ldots,100,101\}$.\footnote{Note that $t(g)=101$ is equivalent to ``never hire any applicant from group $g$.''}  Both AF and SD are interested in the pair of thresholds used by the employer, $r(\text{male})$ and $r(\text{female})$.  This stylized setting allows us to clearly identify \textbf{discrimination} between the two groups: whenever $r(\text{male})\neq r(\text{female})$, the employer would hire applicants from one group with a test score that would \textbf{not} lead to hiring an applicant from the other group.  Thus, if $r(\text{male})< r(\text{female})$, then the employer applies a ``less demanding threshold'' when deciding whether to hire a male applicant than when deciding whether to hire a female applicant.

The principal distinction between AF and SD is essentially whose welfare---the applicants' or the employer's---the algorithm should prioritize, and the effect of this prioritization on outcomes.  The two approaches can be compared with respect to the types of questions they tend to tackle as follows.
\begin{itemize}
\item Studies of \textbf{Algorithmic Fairness} tend to focus on questions like:
	\begin{enumerate}
	\item How do the thresholds affect the applicants' welfares?
	\item What does it mean to treat applicants from both groups of applicants \textbf{fairly}?
	\item Which pair(s) of thresholds (if any) treat both groups of applicants fairly?
	\end{enumerate}
\item Studies of \textbf{Statistical Discrimination} tend to focus on questions like:
	\begin{enumerate}
	\item How do the thresholds affect the employer's welfare?
	\item Which pair(s) of thresholds maximize the employer's welfare?
	\item What factors might justify the employer using different thresholds for the two groups?
	\item How do these thresholds affect individual and group behavior?
	\end{enumerate}
\end{itemize}
We now discuss a recent study of algorithmic fairness in the real world---traffic cameras in Chicago---as a way of illustrating the difference between these related analytical frameworks.

\paragraph{Traffic Cameras, Fairness, \& Discrimination.}  A recent concrete example of AF is provided by a recent ProPublica study of traffic cameras in Chicago.\footnote{Hopkins, Emily and Sanchez, Melissa, ``Chicago’s `Race-Neutral' Traffic Cameras Ticket Black and Latino Drivers the Most,'' \textit{ProPublica}, January 5, 2022: https://www.propublica.org/article/chicagos-race-neutral-traffic-cameras-ticket-black-and-latino-drivers-the-most.}  The study found that, in Chicago in 2020, ``the ticketing rate for households in majority-Black ZIP codes jumped to more than three times that of households in majority-white areas. For households in majority-Hispanic ZIP codes, there was an increase, but it was much smaller.''  An AF perspective on this situation essentially asks why this disparity emerges and, more provocatively, how one might reduce or eliminate it.  This perspective is particularly helpful in this type of setting because, while this disparity has widened over the two decades since the cameras were introduced in Chicago, there is little reason to suspect that traffic cameras themselves are distinguishing between drivers based on their race or home neighborhood, per se.  In this specific case, this perspective allows one to see that the disparity is at least arguably due to speed limits and driving conditions being distributed in a ``non-race blind'' fashion across Chicago.  

\paragraph{Enforcement, Algorithmic Fairness, \& Welfare.} The analysis also demonstrates the complexities of AF in the real world.  Chicago Mayor Lori Lightfoot's administration described traffic cameras as a tool to ``a tool in the toolkit to help alleviate'' traffic fatalities and, from an empirical standpoint, Black Chicagoans were twice as likely to die in a traffic accident as white Chicagoans in 2017.  Accordingly, Black Chicagoans are differentially treated by both traffic accidents \textit{and} traffic tickets.  As we return to below in Section \ref{Sec:Examples}, this is a classic conundrum in AF settings --- an algorithm intended to protect certain individuals might be more error-prone excatly when it is ``interacting with,'' or ``treating,'' those individuals.

\paragraph{Motives, Outcomes, \& Statistical Discrimination.} From a SD standpoint, on the other hand, one might ask \textit{why} Chicago is using traffic cameras, in spite of the clear racial disparity in which citizens receive tickets.  As the ProPublica article describes, Chicago Mayor Lori Lightfoot---despite arguing as a candidate that Chicago should reduce its dependence on traffic tickets and other fines as a revenue source---proposed lowering the minimum speed at which a speeding ticket would be issued.  This proposal, which was adopted by the Chicago City Council in 2021, prompted some to question how much Mayor Lightfoot cared about racial disparities, as opposed to the City of Chicago's serious structural deficit.  The question of ``is Mayor Lightfoot more interested in racial equality or city revenue?'' is directly analogous to the seminal question in statistical discrimination, ``is that employer simply maximizing profits or are they racist?''  Accordingly, in the SD literature, theoretical work (some of which is summarized in Section \ref{Sec:Statistical}) has established a partial typology of ``non-racist'' explanations for ``apparently racist'' behavior.

In terms of the traffic camera example, we can also distinguish the AF and SD viewpoints as 
\begin{itemize}
\item \textbf{Algorithmic Fairness}: Can we make Chicago's traffic enforcement more fair?  If so, how?
\item \textbf{Statistical Discrimination}: Why did Chicago use an unfair traffic enforcement algorithm?
\end{itemize}
We believe both of these questions are important.  Unfortunately, answering one typically requires some assumptions that are either ``outside'' or, in some circumstances, at odds with the intents and purposes of the other question.  SD has a longer intellectual history and, partially as a result, provides an excellent example of such a conflict.  Specifically, as we discuss in Section \ref{Sec:Statistical}, one strand of work assumes that applicants from the male and female groups have different rates of qualification for exogenous reasons (such as historical discrimination, differences in income, \textit{etc.})\footnote{These are referred to as \textit{Phelpsian models of statistical discrimination}, as discussed in Section \ref{Sec:PhelpsianModels}.} whereas another important strand of work explicitly presumes that the two groups are identical in \textit{ex ante} terms and exhibit different behaviors due to differences in actors' \textit{beliefs} about each other.\footnote{These are referred to as \textit{Arrovian models of statistical discrimination}, as discussed in Section \ref{Sec:ArrovianModels}.}

\subsection{Rationality Versus Fairness\label{Sec:RationalityAndFairness}}

A theme running throughout this article is that a key contrast between the AF \& SD approaches revolves around the question of \textbf{rationality} or, in slightly different terms, \textbf{efficiency}.  Many SD theories are focused on how the pursuit of efficiency (\emph{e.g.}, by an employer, job applicant, government, or other individuals) can generate behavior that is discriminatory.  On the other hand, AF is less concerned with efficiency (partly because the framework does not presume anything about individuals' motives/goals).  

This disjuncture is itself informative: it partially isolates a fundamental tension between rationality/efficiency and fairness in many settings.  By taking this contrast seriously, one can leverage the various layers of conflict between various notions of fairness to link them with various actors' instrumental motivations.  Taken as a whole, the SD literature is less focused on any specific notion of discrimination or fairness that one is concerned with and instead focuses on how the pursuit of instrumental goals might have spillover effects on fairness, broadly written.  Similarly, the younger AF literature is less concerned with any specific notion of efficiency, implying that its conclusions are robust to the specific goals of individuals within the algorithmic process in question.

\subsection{Outline of the Article} 

We structure the article as follows.  We first describe a few real-world examples where issues of AF and SD loom large.  We will return to one of these examples in later sections in the hopes of making the various fairness concepts (and the stakes associated with them) more concrete.  In Section \ref{Sec:Theoretical} we introduce notation and in Section \ref{Sec:Fairness} we provide a brief introduction to some of the more well-known concepts in AF.  Section \ref{Sec:Impossibility} describes how well-known concepts in AF may be at odds with each other.  Section \ref{Sec:Statistical} lays out a simple model of SD in order to illustrate some connections with questions of AF.

\section{Fairness \& Discrimination in Employment\label{Sec:Examples}}

The AF and SD literatures are both relevant to a wide array of common economic, social, and political decisions.\footnote{Some other common decisions in which algorithms, and their fairness, have attracted attention recently include college admissions, lending (\textit{e.g.}, \cite{MunnellTootellBrowneMcEneaney96}, \cite{Ladd98}), housing (\textit{e.g.}, \cite{FoggoVillasenor21}),  criminal sentencing (\textit{e.g.}, \cite{Washington18}), and advertising (\cite{MillerHosanagar19}).}    In this section we briefly explore these connections within the most widely studied setting: employment decisions.

Before discussing these, however, it is important to note that AF is relevant for \textit{any} decision-making process in which decisions depend on individual- and/or group-level traits.  Accordingly, the questions examined in this literature are not tied to the use of an explicitly described algorithm (much less a complicated one).  Similarly, the issues raised here are relevant to SD in a wide variety of situations, including many in which the term ``discrimination'' is not widely used.

\paragraph{Employment.}  A firm is faced with the possibility of hiring an individual.  The applicant has two observable traits: gender (male or female) and a test result.  In addition, the applicant has an unobserved trait (skilled or not).  Though the firm can't observe the applicant's skill it wants to hire the individual if and only if he or she is skilled.

A key question for the employer, of course, is how to figure out whether the applicant is skilled or not.  Skill may be correlated with either or both gender and education.  In such cases, it is permissible for the firm to base its hiring decision on education, but not on gender.  Hiring on the basis of an observed trait is known as \textit{disparate treatment}.  Disparate treatment on the basis of education is legal, but on the basis of gender is illegal.  

\textit{Disparate impact} occurs when the outcomes experienced by one gender are better than the other.\footnote{In practice, the difference in outcomes must exceed some positive threshold, the level of which varies across different contexts.}  For example, if the employer hires 50\% of male applicants, but only 10\% of female applicants, then its hiring process exhibits disparate impact, regardless of whether the process also exhibits disparate treatment.  In practice, illegal disparate impact in employment typically occurs when the employer uses one or more indicators that are not a ``reasonable measure of job performance.'' 

Disparate impact is the primary focus of AF in hiring and promotion decisions.  Conversely,  the literature on SD---with its focus on the equilibrium actions of goal-oriented agents---is more often concerned with questions of disparate treatment.  Disparate treatment is both more straightforward and illegal, \textit{per se}, whereas disparate impact is not illegal if it is caused by some factor that is sufficiently correlated with job performance.\footnote{The question of whether intent is relevant to establishing disparate impact is a complicated one (\textit{Texas Dept. of Housing and Community Affairs v. Inclusive Communities Project, Inc.}, 576 U.S. 519 (2015)).}   Moreover, remedying disparate impact may necessitate disparate treatment. Two central conclusions are:
\begin{enumerate}
    \item Mitigating disparate treatment can exacerbate disparate impact (and vice-versa), and
    \item It is generally impossible to eliminate both disparate treatment and disparate impact.
\end{enumerate}

\paragraph{The Potential Conflict Between Disparate Treatment and Disparate Impact.} The two conflicts between minimmization of disparate treatment and disparate impact can be seen in practice in \textit{Ricci v. DeStefano}, 557 U.S. 557 (2009).  This case, brought under the Civil Rights Act of 1964, focused on the use of a exam for promotion within the City of New Haven, CT's Fire Department.  Twenty firefighters---19 of whom were white and one of whom was Hispanic---passed the exam, but were not promoted because no Black firefighters passed the exam. The City of New Haven ignored the test results because they worried about a disparate impact claim with respect to the exam's outcomes.  The twenty non-Black firefighters argued that ignoring the exam results constituted disparate treatment under Title VII of the Civil Rights Act of 1964.

The Supreme Court ruled that the City of New Haven's decision to ignore the test results was illegal because the Court did not agree that New Haven would have been subject to a charge of disparate impact if, after administering the test, the fire department had failed to promote any Black firefighters.  Indeed, the legal reasoning's complexity is at least in part due to the fact that minimizing disparate treatment while also minimizing disparate impact was impossible in this specific case.  Perhaps obviously, this type of situation is not uncommon.  For the remainder of this article we simplify matters and use the term ``discrimination'' to refer to either disparate treatment or disparate impact.

\section{Theoretical Fundamentals \label{Sec:Theoretical}}

We now define the standard primitives joining the AF and SD frameworks. For simplicity, we focus on a common setting for this type of work, motivated by the employment setting briefly discussed above.  In this setting, there is an \textbf{employer} and a pool of \textbf{applicants}.  The employer will decide whether to hire each applicant on the basis of the applicant's observable characteristics (\textit{e.g.}, education, prior work history, entrance exams, credit history, \textit{etc}.). 

Each applicant has a single, \textit{unobserved} characteristic that is of interest to the decision-maker (\textit{e.g.}, is the individual ``qualified'' for the job or not). For any applicant, the ``hiring algorithm'' (which might ``represent a strategic employer'' or not) makes a binary choice (\textit{e.g.} to hire or not). Hiring a qualified individual or not hiring an unqualified individual are each considered a \textit{success}, while hiring an unqualified applicant or not hiring a qualified applicant are each considered \textit{failures} of the algorithm.  

Thus, the ``employer's objective'' for the algorithm is essentially to correctly predict whether each applicant is qualified or not.   While we attempt to discuss this material in as transparent a fashion as possible, a little notation will greatly aid our presentation and, more importantly, comparability of our arguments with those in the AF and SD literatures.\footnote{We generally follow the notation and terminology of \cite{MitchellPotashBarocasDAmourLum21}.}

\subsection{Basic Building Blocks}

As described above, we consider a setting with an employer, $E$, and a pool of applicants, $N=\{1,2,\ldots,n\}$.  Each applicant $i\in N$ is described by the following:
\begin{enumerate}
    \item A profile of \textit{permissible} traits, $x_i=(x_i^1,\ldots, x_i^m)$, \\
		\textbf{Examples}: Education, technical skills, test scores, credit history
    \item A profile of \textit{sensitive} traits, $a_i=(a_i^{m+1},\ldots,a_i^M)$,\\
		\textbf{Examples}: Gender, race, ethnicity, marital status
    \item An \textit{outcome}, $y_i \in\{0, 1\}$,\\
		\textbf{Examples}: Qualification for the job, profitability of investment, efficacy of treatment
    \item A \textit{decision}, $\delta_i \in \{0, 1\}$. \\
		\textbf{Examples}: Did $i$ get the job?  Did $i$ get admitted?  Did $i$ get the loan?
\end{enumerate}
Unless otherwise stated, $x_i$, $a_i$, 
and $\delta_i$ are assumed to be observable to the employer, but $y_i$ is not.  We briefly describe each of these terms below.  Note that for the moment we assume that traits and outcomes are exogenous and fixed.  We will relax this assumption later when describing paths for future research, by allowing individuals to potentially exert some control over both.

\paragraph{Traits.} We will write $v_i=(a_i,x_i)$ to denote individual $i$'s sensitive and permissible traits and $V$ to denote the set of all profiles of individual traits.  The key distinction between the types of traits is that it is ``okay'' to discriminate between individuals on the basis of differences in their permissible traits, but potentially ``not okay'' to discriminate between them on the basis of one or more sensitive traits.  In the employment example, the applicant's education may be a permissible trait, while his or her gender may be a sensitive trait.  

\paragraph{Outcomes.} An individual's outcome is a binary measure of a characteristic that is of interest to the employer.  In the hiring context, $y_i$ might represent whether the individual would succeed in the job if hired.
In the education context, $y_i$ might represent whether $i$ will graduate from college.  In the lending example, $y_i$ might represent whether $i$ will repay the loan if it is granted to $i$. 


\paragraph{Decisions.}  Finally, the goal of the algorithm is to ultimately generate decision(s) that will possibly affect the individuals in question. These decisions are a function of each individual's traits, with $\delta_i\in\{0, 1\}$ being the \textit{decision} regarding person $i$.  In our employment example, $\delta_i=1$ could represent the decision to hire individual $i$, and $\delta_i=0$ would be the decision to not hire $i$.  In light of these definitions, our \textit{algorithm} is the collection of decisions, $\delta$.

\subsection{The Confusion Matrix}

We conclude our discussion of ``theoretical fundamentals" with an illustration of a \textbf{confusion matrix} for this setting of binary outcomes and binary decisions.  The simplified setting is useful because it focuses attention on exactly two kinds of mistakes: Type I and Type II errors.  This makes it easy to consider and compare fairness goals that focus on the algorithm's relative \textbf{predictive performance} across sensitive traits, as discussed below in Section \ref{Sec:MeasuresOfPredictivePerformance}.  That said, this convenience is not without some loss of generality because the 2x2 case sets aside a range of considerations that are potentially important, such as the possibility of ``small" versus ``large" misclassfications by a decision-maker or algorithm.

 \begin{table}[h!]
 \scriptsize
 \centering
\begin{tabular}{|c|c|c|c|}  
\cline{2-3} 
\multicolumn{1}{c|}{}   & \multicolumn{2}{|c|}{\textbf{Predicted Outcome}}         & \multicolumn{1}{c}{} \\
\cline{1-3}
\textbf{True Outcome}      &  Positive ($\delta_i=1$)	& Negative ($\delta_i=0$)   & \multicolumn{1}{c}{} \\
\hline
& & & \cellcolor[gray]{0.9}   \\
Positive ($y_i=1$) 
& \begin{tabular}{c}
     True Positives (TP)\\
\end{tabular}
& \begin{tabular}{c}
     False Negatives (FN)\\
\end{tabular}
& \cellcolor[gray]{0.9}  \begin{tabular}{c}
     \textbf{True Positive}\\
     \textbf{Rate (TPR)}\\
     $\frac{TP}{TP+FN}$
\end{tabular}\\
& & & \cellcolor[gray]{0.9}   \\
\hline
& & & \cellcolor[gray]{0.9}   \\
Negative ($y_i=0$)
& \begin{tabular}{c}
     False Positives (FP)\\
\end{tabular}
& \begin{tabular}{c}
     True Negatives (TN)\\
\end{tabular}& \cellcolor[gray]{0.9}   \begin{tabular}{c}
     \textbf{True Negative}\\
     \textbf{Rate (TNR)}\\
     $\frac{TN}{FP+TN}$
\end{tabular}\\
& & & \cellcolor[gray]{0.9}   \\
\hline
\multicolumn{1}{c|}{} & \multicolumn{1}{|c|}{\cellcolor[gray]{0.9}}   & \cellcolor[gray]{0.9}  & \multicolumn{1}{c}{} \\
\multicolumn{1}{c|}{} & \multicolumn{1}{|c|}{\cellcolor[gray]{0.9}   \begin{tabular}{c}
     \textbf{Positive Predictive}\\
     \textbf{Value (PPV)} \\
     $\frac{TP}{TP+FP}$
\end{tabular}}
& \multicolumn{1}{c|}{\cellcolor[gray]{0.9}   \begin{tabular}{c}
     \textbf{Negative Predictive}\\
     \textbf{Value (NPV)}\\
     $\frac{TN}{TN+FN}$
\end{tabular} }& \multicolumn{1}{c}{}\\
\multicolumn{1}{c|}{}& \cellcolor[gray]{0.9}  & \cellcolor[gray]{0.9}  & \multicolumn{1}{c}{}\\
\cline{2-3}
\end{tabular}
 \caption{Predictive Performance with Binary Outcomes and Decisions \label{confusion}}
\end{table}

\normalsize

Table \ref{confusion} describes several well-known measures of the predictive performance of an algorithm in this 2x2 case.  These measures are used to inform a number of the fairness goals that we will shortly define in Section \ref{Sec:Fairness}. Before proceeding, we will simply note that different stakeholders may prioritize these measures very differently (\cite{Narayanan18}).  Suppose, for example, that outcomes represent whether a worker is qualified ($y_i=1$) or not ($y_i=0$), and decisions represent whether the worker is hired ($\delta_i=1$) or not ($\delta_i=0$).  A qualified applicant may care most about the \textit{true positive rate} of the hiring algorithm, or the probability that they get the job.  A firm may care most about the \textit{positive predictive value} of the hiring algorithm, or the probability that a worker that is hired is actually qualified for the job.  In general, the relevant actors may care most about measures that condition on their own traits or decisions (outcome $y_i$ for the worker;  decision $\delta_i$ for the employer).

\section{Algorithmic Fairness Goals\label{Sec:Fairness}}

\cite{CorbettDaviesGoel18} define the following two (non-exclusive) families of fairness goals that can be thought of as ``fairness from the data alone."  As \cite{MitchellPotashBarocasDAmourLum21} observe, these definitions equate fairness with equalities that can be derived solely from the distribution of traits, outcomes, and decisions.

\begin{enumerate}
    \item {\bf Anti-classification.} Sensitive traits are not directly used to make decisions.
    \item {\bf Classification Parity.} Predictive performance is independent of sensitive traits.
\end{enumerate}
\cite{CorbettDaviesGoel18} illustrate some pathological consequences that can follow from achieving anti-classification and/or classification parity, as well as a general incompatibility between measures of classification parity.  We now define and briefly discuss these categories of fairness goals.

\subsection{Anti-classification} 

In many contexts it is inadmissible or illegal to condition one's decision on an individual's sensitive traits.  Doing so constitutes disparate treatment under US federal employment law.  There has consequently been a clear focus on ensuring that sensitive traits do not factor into how an algorithm or employer evaluates individuals.  An algorithm satisfies anti-classification if two individuals with the same permissible traits receive the same decision, or:
    \[
    x_i=x_j \Rightarrow \delta_i = \delta_j.
    \]
Anti-classification restricts the information that decisions can be responsive to. In this sense,  of the three categories of fairness goals, it is the most clearly associated with \textbf{process}: what factors can directly affect the algorithm's decision for any given individual?  In addition, it is also trivially satisfiable.  For example, anti-classification is satisfied simply by having the algorithm assign every individual the same decision ($\delta_i=\delta_j$ for all $i, j$).  

In spite of this pathology, the criterion has important appeal as well.  For example, as a direct reflection of a desire to not discriminate against individuals \emph{solely} on the basis of a sensitive trait such as race or gender, it represents a minimal but clear attempt to reduce or eliminate \emph{taste-based} discrimination.  This is a large part of the reason that anti-classification is the focus of (and defends against) disparate treatment claims.  An important foundational implication of both AF and SD theories is that anti-classification alone will not only not necessarily eliminate discrimination in practice, because of the possibility of disparate impact, but that satisfying it can in some situations \emph{exacerbate} discrimination.

\subsection{Measures of Predictive Performance \label{Sec:MeasuresOfPredictivePerformance}} 

\textit{Predictive performance} is a somewhat loosely defined concept describing various approaches to evaluating how well an algorithm ``works.''  There are many measures of predictive performance: which measure(s) are appropriate depend on the goals/outputs of the model in question.  A classic distinction, for example, divides prediction problems into \textbf{classification}---in which the model is attempting to assign individuals/data points to discrete (often ``unordered'') categories---or \textbf{regression} problems, in which the model is associating each individual/data point with a real number.  Given our central focus in this article, we consider only classification measures of predictive performance. 

Any measure of predictive performance can be used to define a related notion of fairness, often referred to as a form of \emph{parity} or \emph{balance}.  Satisfaction of such parity/balance requirements typically requires that the chosen measure(s) of predictive performance be independent of an individual's sensitive traits. 
Recently, scholars have noticed that many of these measures of parity/balance can be inconsistent with each other.\footnote{For example, \cite{KleinbergMullainathanRaghavan16}, \cite{Chouldechova17}, and \cite{BerkHeidariJabbariKearnsRoth18}.}  These \textbf{impossibility results} imply that, in many cases, \emph{any} algorithm must be ``unfair'' by at least one measure of predictive performance,

\paragraph{Comparing Predictive Performance Across Sensitive Traits.}  The predictive performance of an algorithm in the 2x2 baseline case is described by its confusion matrix (Table~\ref{confusion}, above).  The table contains a wide array of information.  A key point is the fact that the positive and negative predictive values, PPV and NPV, are conditioned on the decision, $\delta_i$, whereas the true positive and true negative rates, TPR and TNR, are conditioned on the outcome, $y_i$.  They may look similar but, as is often the case with conditional probabilities, the relationship between the two can be counterintuitive and depends upon the latent distribution of outcomes, $y_i$.  A key finding in this literature is that different fairness goals will tend to suggest different algorithms whenever the distribution of outcomes, $y_i$, depends non-trivially on the individuals' sensitive traits, $a_i$.  We now discuss three particularly well-known measures of classification parity: \textit{predictive parity}, \textit{error rate balance}, and \textit{demographic parity}.\\

\noindent\textbf{\textsc{Predictive Parity.}} Predictive parity captures the idea that, conditional on the decision $\delta$, individuals with different sensitive traits should be equally likely to have the same outcome $y_i$.  This boils down to equality of positive predictive values (PPV) across groups, or negative predictive values (NPV), or both. Formally, predictive parity is defined as follows.
\begin{definition}[Predictive Parity]
\label{Def:PredictiveParity}
An algorithm $\delta$ satisfies
\begin{enumerate}
    \item {\bf Positive predictive parity} if each pair of groups, $a_i, a_j$, have the same positive predictive value:
\begin{equation}
\label{Eq:PositivePredictiveParity}
PPV(a_i) \equiv \Pr[y_i=1 \mid \delta_i=1, a_i]=\Pr[y_i=1 \mid \delta_i=1, a_j] \equiv PPV(a_j).
\end{equation}
    \item {\bf Negative predictive parity} if each pair of groups, $a_i, a_j$, have the same negative predictive value: 
\[
NPV(a_i) \equiv \Pr[y_i=0 \mid \delta_i=0,  a_i]=\Pr[y_i=0 \mid \delta_i=0, a_j] \equiv NPV(a_j).
\]
\end{enumerate}
Finally, $\delta$ satisfies {\bf predictive parity} if $\delta$ satisfies both positive \&  negative predictive parity.
\end{definition}
\mbox{}

\noindent\textbf{\textsc{Error Rate Balance.}} Error rate balance compares true and false positive rates and requires that individuals differing only with respect to sensitive traits are equally likely to be mis-classified by the algorithm.  Formally, error rate balance is defined as follows.  

\begin{definition}[Error Rate Balance]
\label{Def:ErrorRateBalance}
An algorithm $\delta$ satisfies {\bf positive error balance} if each pair of groups, $a_i, a_j$, have the same true positive rate:
\[
TPR(a_i) \equiv \Pr[\delta_i=1 \mid y_i=1, a_i]=\Pr[\delta_i=1 \mid y_i=1, a_j] \equiv TPR(a_j).
\]
Similarly, an algorithm $\delta$ satisfies {\bf negative error balance} if each pair of groups, $a_i, a_j$, have the same true negative rate:
\[
TNR(a_i) \equiv \Pr[\delta_i=0 \mid y_i=0,  a_i]=\Pr[\delta_i=0 \mid y_i=0,  a_j] \equiv TNR(a_j).
\]
Finally, $\delta$ satisfies {\bf error rate balance} (ERB) if $\delta$ satisfies both positive \&  negative error balance.
\end{definition}
\mbox{}\\

\noindent\textbf{\textsc{Demographic Parity.}} Demographic parity (sometimes referred to as statistical parity or group fairness) is a widely employed fairness criterion.  Substantively, demographic parity is satisfied when sensitive traits do not affect the distribution of decisions for a randomly drawn individual.  Demographic parity has several virtues: it is transparent and simple to calculate, and it cleanly captures one classic notion of equality of outcomes. Of course, one downside of demographic parity is that it can be satisfied by an algorithm that doesn't respond to permissible traits, or any data at all.  Formally, demographic parity is defined as follows.
\begin{definition}
\label{Def:DemographicParity}
An algorithm $\delta$ satisfies {\bf demographic parity}  if, for any pair of profiles of sensitive traits, $a_i, a_j$, 
\[
\Pr[\delta_i \mid a_i] = \Pr[\delta_i \mid a_j].
\]
\end{definition}

\section{The Inherent Conflict Between Fairness Notions \label{Sec:Impossibility}}

With several notions of fairness at hand, one might naturally ask under what conditions they will be in agreement.  \label{Pg:ImpossibilityResults} For better or worse, recent theoretical work demonstrates that predictive parity and error rate balance are generally not consistent with each other. 
\begin{theorem}[\cite{KleinbergMullainathanRaghavan16}]
\label{Th:Impossibility}
If an algorithm satisfies Predictive Parity and Error Rate Balance, then one or both of the following must be satisfied:
\begin{eqnarray}
\Pr[y_i=1\mid a_i,x_i] & \in & \{0,1\} \text{ for all } x_i, a_i \;\;\; \text{\textbf{Perfect Predictor}}, \label{Eq:PerfectPredictor}\\
\Pr[y_i=1\mid a_i] & = & \Pr[y_i=1\mid a'_i] \text{ for all } a_i,a'_i \;\;\; \text{\textbf{Equal Base Rates}}. \label{Eq:EqualBaseRates}
\end{eqnarray}
\end{theorem}
Theorem \ref{Th:Impossibility} establishes that simultaneously satisfying predictive parity and error rate balance is possible \emph{only if} (1) individual outcomes ($y_i$) can be predicted \emph{perfectly} (``perfect predictor,'' Equation \eqref{Eq:PerfectPredictor}) and/or (2) individual outcomes are uncorrelated with sensitive traits (``equal base rates,'' Equation \eqref{Eq:EqualBaseRates}).  Note that if perfect predictor is satisfied, then there is no uncertainty in the employer's classification problem, implying that ``the optimal algorithm'' is trivial.  If, in addition or on the other hand, equal base rates is satisfied, then the two groups are statistically indistinguishable with respect to outcomes.

\paragraph{The Implications of Impossibility.} At this point, it is worthwhile to briefly think about what impossibility results tell us.\footnote{For a more general take on what impossibility theorems do (and do not) imply, see \cite{PattyPenn14,PattyPenn18}.}  The basic idea of an impossibility result is as follows: ``nothing can satisfy criteria $A$ and $B$ simultaneously unless condition $C$ is satisfied.''  In this terminology, $A$ and $B$ are typically each desiderata that we hope a decision or measure might satisfy, and $C$ is a (typically) ``special'' set of circumstances delimiting the impossibility claim.  In Theorem \ref{Th:Impossibility}, criterion $A$ is ``satisfies PPV,'' criterion $B$ is ``satisfies error rate balance,'' and condition $C$ is ``perfect predictor or equal base rates'' is/are satisfied.  A common misundertanding of such results (such as Arrow's Theorem \cite{Arrow51}, the Gibbard-Satterthwaite theorem (\cite{Gibbard73}, \cite{Satterthwaite75}, \cite{PennPattyGailmard11}), or Sen's ``Paretian Liberal'' (\cite{Sen70})) is that they describe situations in which it is \emph{very difficult} (or, perhaps, it is ``rarely possible'') to satisfy criteria $A$ and $B$ if condition $C$ is not satisfied.  This is a fundamental and important misunderstanding: impossibility theorems tell us that unless condition $C$ is satisfied, criteria $A$ and $B$ \textbf{cannot} be satisfied simultaneously.

\paragraph{Incompatibility of Predictive Parity and Error Rate Balance.}  Viewed from a different perspective, Theorem \ref{Th:Impossibility} establishes that the notions of predictive parity and error rate balance are not only generally incompatible, but also intertwined.  This is intuitive in some ways: they consider two different, but related, conditional probabilities.  Predictive parity concerns the fairness of the algorithm with respect to applicants' outcomes \emph{conditional on the algorithm's decision}. Error rate balance concerns  the fairness of the algorithm with respect to the algorithm's decisions \emph{conditional on the applicants' outcomes}.  One reason that Theorem \ref{Th:Impossibility} is so surprising at first essentially flows from a ``common'' mistake with respect to Bayes's rule:
\begin{eqnarray}
\hline
& & \nonumber \\
PPV & = & \frac{\Pr[y_i=1\text{ \& } \delta_i=1]}{\Pr[\delta_i=1]} = \frac{\Pr[y_i=1\text{ \& } \delta_i=1]}{\underbrace{\Pr[\delta_i=1\text{ \& } y_i=1]}_{\text{\textbf{True Positive}}}+\underbrace{\Pr[\delta_i=1\text{ \& } y_i=0]}_{\text{\textbf{False Positive}}}},\label{Eq:PPVFPR}\\
TPR & = & \frac{\Pr[y_i=1\text{ \& } \delta_i=1]}{\Pr[y_i=1]} = \frac{\Pr[y_i=1\text{ \& } \delta_i=1]}{\underbrace{\Pr[y_i=1\text{ \& } \delta_i=1]}_{\text{\textbf{True Positive}}}+\underbrace{\Pr[y_i=1\text{ \& } \delta_i=0]}_{\text{\textbf{False Negative}}}}, \label{Eq:TPRFNR}\\
 & & \nonumber \\
\hline
 & & \nonumber \\
NPV & = & \frac{\Pr[y_i=0\text{ \& } \delta_i=0]}{\Pr[\delta_i=0]} = \frac{\Pr[y_i=0\text{ \& } \delta_i=0]}{\underbrace{\Pr[\delta_i=0\text{ \& } y_i=0]}_{\text{\textbf{True Negative}}}+\underbrace{\Pr[\delta_i=0\text{ \& } y_i=1]}_{\text{\textbf{False Negative}}}},\label{Eq:NPV}\\
TNR & = & \frac{\Pr[y_i=0\text{ \& } \delta_i=0]}{\Pr[y_i=0]} = \frac{\Pr[y_i=0\text{ \& } \delta_i=0]}{\underbrace{\Pr[y_i=0\text{ \& } \delta_i=0]}_{\text{\textbf{True Negative}}}+\underbrace{\Pr[y_i=0\text{ \& } \delta_i=1]}_{\text{\textbf{False Positive}}}}, \label{Eq:TNR}\\
& & \nonumber \\
\hline \nonumber 
\end{eqnarray}
Equations \eqref{Eq:PPVFPR} -- \eqref{Eq:TNR} identify the interlocking connections between PPV, NPV, TPR, and TNR.  Comparing the pairs of labels in the denominators of the right hand sides of (for example) \eqref{Eq:PPVFPR} \& \eqref{Eq:TPRFNR}, notice that they have one additive term in common (the total frequency of true positives), but differ with respect to the second term of the corresponding sums (total frequency of false positives for PPV and total frequency of false negatives for TPR).  A fundamental feature of (``most'') non-trivial algorithmic problems (captured by the perfect predictor condition, \eqref{Eq:PerfectPredictor}, in Theorem \ref{Th:Impossibility}) is that these ``unmatched'' terms of the sums in the two denominators are each non-zero.  A foundational fact for Theorem \ref{Th:Impossibility} and other impossibility theorems about measures of predictive performance is that the three terms contained in one or both of the denominators on the right hand sides of \eqref{Eq:PPVFPR} \& \eqref{Eq:TPRFNR} represent all but one of four nonnegative numbers that must sum to one.  Accordingly, \textit{once the denominators of \eqref{Eq:PPVFPR} \& \eqref{Eq:TPRFNR} are fully specified, the denominators of both \eqref{Eq:NPV} and \eqref{Eq:TNR} are fully specified as well.}

\paragraph{Predictive Parity \& Error Rate Balance in Conflict, in Practice.} COMPAS (Correctional Offender Management Profiling for Alternative Sanctions) is a commercially marketed tool designed to assess the \textbf{recidivism risk} of criminal defendants.  It has been shown that COMPAS violates error rate balance: an investigation published in ProPublica by \cite{Angwin2016} showed that, while the tool satisfies positive predictive parity, it violates error rate balance in that Black defendants were twice as likely to be mis-classified as ``high risk'' as white defendants, and white defendants were twice as likely to be mis-classified as ``low risk'' as Black defendants.  \cite{Chouldechova17} showed, via an admirably simple identity, that positive predictive parity is incompatible with error rate balance unless there are equal base rates (outcomes are independent of sensitive traits, which, in the COMPAS debate, would necessitate equal rates of recidivism for Blacks and whites), or the algorithm disallows certain types of errors (specifically false positives).
This result, proved independently of the impossibility theorem of \cite{KleinbergMullainathanRaghavan16}, demonstrates that satisfaction of positive predictive parity generally requires that error rate balance be violated.

\paragraph{Positive Predictive Parity in Hiring.}  In terms of employment decisions, positive predictive parity requires that the outcomes among those who are hired are distributed identically across individuals possessing different sensitive traits.  Considering gender as the sensitive trait, for example, positive predictive parity in employment requires that men who are hired are equally as likely to be skilled as women who are hired. The distribution of outcomes among the hired will typically tend to differ from those not hired, but the distribution of outcomes in either case should not depend on the individuals' sensitive traits.  (Conversely, negative predictive parity requires that the distribution of skills among those \textit{not} hired be identical for men and women.)  If positive predictive parity is violated, then the employer should differentiate between hired individuals on the basis of gender for subsequent promotion and retention decisions (see \cite{Fryer07} for an excellent discussion of this issue that is directly relevant to the focus of our article).

Error rate balance focuses on  fairness conditional upon individual outcome, rather than algorithm's decision.  Just as predictive parity tends to weigh more heavily in the decision-maker's subsequent decision-making, error rate balance may tend to weigh more heavily in applicants'  decision-making, particularly when outcomes are endogenous and/or the algorithm is sensitive to individual outcomes.\footnote{We consider both of these possibilities in Section \ref{Sec:Statistical}, below.}  This is because the error rates associated with an algorithm will affect the marginal costs and benefits of investing in a positive outcome to an applicant motivated by a desire for a positive decision from the algorithm. 

To see this, consider an employment setting in which the individual outcome corresponds to the skill of the applicant, and this skill is the product of costly investment by the applicant.  Violation of error rate balance may lead to a correlation between an individual's sensitive traits and his or her (1) application for the job and/or (2) investment in skill aquisition.  To the degree that the employer is truly indifferent about sensitive traits, per se, this may not have an impact on the employer.

\subsection{Demographic Parity: Motivations and Challenges}

As mentioned above, demographic parity is appealing for a variety of reasons, including its simplicity and transparency.  However, it is a ``group-level'' measure of equality, meaning that it does not account for the possibility that the distribution of permissible traits and/or outcomes may differ across (sensitive trait-defined) groups.  For example, if individuals with one value of a sensitive trait are more likely than those with a different value of that trait to have an outcome of $y_i=1$, then demographic parity is not necessarily a reasonable fairness goal from an equity and/or social welfare standpoint.

Less provocatively, it might be the case that outcomes are independent of sensitive traits, but correlated with one or more permissible traits that are themselves correlated with sensitive traits.  For example, gender might not be correlated with skill once one ``controls for'' permissible traits such as education, physical ability, etc., but correlated with skill if one ignores these factors.  A straightforward extension of demographic parity to allow for such inter-group differences with respect to permissible traits can be defined as follows:

\begin{definition}
\label{Def:ConditionalDemographicParity}
An algorithm $\delta$ satisfies {\bf conditional demographic parity}  if, for any pair of profiles of sensitive traits, $a_i, a_j$, 
\[
x_i=x_j \Rightarrow \Pr[\delta_i \mid a_i,x_i] = \Pr[\delta_i \mid a_j,x_j].
\]
\end{definition}  This generalized notion of demographic parity is an ``individual-level'' measure of equality, because it accounts for differences between individuals within a given group.  Referred to by several names, including \textit{conditional demographic parity} (\cite{MitchellPotashBarocasDAmourLum21}), \textit{unawareness} (\cite{KusnerLoftusRussellSilva17}), or \textit{treatment parity} (\cite{LiptonChouldechovaMcAuley18}) these notions are each equivalent to anti-classification in the settings considered here (\cite{MitchellPotashBarocasDAmourLum21}, p.153).  While not our main point in this article, it is useful to notice that these equivalence results provide further evidence of the interconnectedness of many of the fairness notions currently being analyzed and/or used in practice. Somewhat less opaquely, many notions of algorithmic fairness explicitly and/or implicitly combine ``various components of each other.''  When this is done explicitly, we view this aggregation of notions as a constructive and sometimes illuminating exercise.  However, impossibility results like Theorem \ref{Th:Impossibility} imply that things are much murkier when the overlap is less transparent, because one can easily create two or more aggregated notions of fairness that are inconsistent with each other.

In any event, conditional demographic parity has at least one strength relative to absolute demographic parity (it is sensitive to permissible traits) but also at least one weakness (satisfaction of conditional demographic parity \emph{requires} satisfaction of anti-classification).  Accordingly, we set demographic parity to the side for the majority of the discussion below.

\section{Statistical Discrimination\label{Sec:Statistical}}

Models of SD aim to explain unequal outcomes across groups game theoretically, as the product of rational decision-making by individuals. As in the literature on AF, these models typically assume that a decision-maker is imperfectly informed about some important characteristic held by an individual.  If group membership is correlated with the characteristic then it is rational for the decision-maker to use group statistics as a proxy for the unobserved individual-level characteristic. 

\cite{Fang11} provide a comprehensive review of the literature on SD, and note that, broadly speaking, there are two strands of models that correspond to two different explanations for inter-group inequality.  

\paragraph{Phelpsian Models of Statistical Discrimination.} The first class of models discussed by \cite{Fang11} build on work by \cite{Phelps72} and assumes that there may be intrinsic differences between the groups.  These differences may be in outcomes (\textit{e.g.} women are, on average, more skilled than men) or in how information about outcomes is conveyed to the decision-maker (\textit{e.g.} men and women are equally skilled on average, but education is a noisier/less informative signal of skill for men than for women).   In both cases, it can be rational for the decision-maker to treat members of the two groups differently.  That said, the nature of the distinction between the groups (different prevalence of outcomes vs. differently reliable measures of these outcomes) is relevant for the possibility of achieving different fairness goals, as we return to in Section \ref{Sec:PhelpsianModels}, below.  

\paragraph{Arrovian Models of Statistical Discrimination.} The second class of models discussed by \cite{Fang11} build on work by \cite{Arrow73} and assume that the groups are ex-ante identical. These models yield multiple equilibria, with inequality being characterized by coordination on different equilibria by different groups.  The decision-maker ultimately arrives at different beliefs about the two groups, and these beliefs are correct (``self-confirming") and derived in equilibrium.  Importantly, these models assume endogeneity of the outcome variable, $y_i$.  Suppose that $y_i$ corresponds to whether $i$ is skilled or not, but that skill is the product of some costly investment by the worker. If, for example, an employer believes that $y_i=0$ for all individuals in a group then those individuals will have no incentive to invest in becoming skilled, because they have no hope of being hired.  This leads to the employer having correct beliefs about the group, in equilibrium.

In the following sections we present a simple model of SD that can accommodate features of both the Phelpsian and Arrovian frameworks.  We then use this baseline model to examine questions of AF within a straightforward setting of strategic interaction.  We use much of the same notation presented in Section \ref{Sec:Theoretical} in order to draw connections between these two literatures.

\subsection{\label{Sec:Model} A Simple Model of Statistical Discrimination}

Consider a situation in which an employer $E$ is deciding whether to hire a worker $W$.  $W$ is either qualified for the job or not.  $W$'s outcome variable, $y_W\in\{0, 1\}$, is referred to as $W$'s \textit{qualification}.  If $y_W=1$ then $W$ is qualified and if $y_W=0$ then $W$ is not qualified.  It is assumed that $y_W$ is private information, known only by $W$.\footnote{Knowledge of $y_W$ by $W$ is not necessary for our analysis but clarifies some of the main theoretical points.}  

In addition to being either qualified or not, $W$ also has two potentially observable traits.\footnote{Unless stated otherwise, we assume that both traits are observable by the employer.}  Gender, $a_W\in \{m, f\}$ is a sensitive trait.  A test score $x_W\in\{1, 2, 3\}$ is a permissible trait.  The information about $W$ that is available to $E$ is  $v_W\in\{m, f\}\times\{1, 2, 3\}$. 
The fraction of males that are qualified for the job is denoted by $p_m\in [0,1]$ and the fraction of females that are qualified for the job is similarly denoted by $p_f$.  The terms $p_m$ and $p_f$ represent the ``prevalence of qualification" within each respective group, or the base rates of qualification, and $E$ is assumed to have correct beliefs about these values.  

\paragraph{The Employer's Payoffs.} We assume that $E$ receives a payoff of $B>0$ for hiring a qualified worker, a payoff of $\omega<0$ for hiring an unqualified worker, and a payoff equal to 0 for not hiring the worker (regardless of $W$'s qualification).  Note that none of these payoffs depend on the worker's gender, $a_W$.  This implies that any employment discrimination within this setting is \textit{not} taste-based (\cite{Becker71}).

\paragraph{The Employer's Beliefs.} Working within a game theoretic tradition, we assume that $E$ always forms correct beliefs about $W$ qualification, $y_W$, conditional on the information $E$ has (\textit{i.e.}, conditional on $v_W$). These beliefs represent a \textbf{calibrated assessor}, in the language of AF.
Consequently, $E$'s beliefs about $W$'s qualification conditional on observing $W$'s gender and test score will coincide with the probabilities described in Equations \eqref{Eq:EmployerBeliefsGenderBlindTesting} or \eqref{Eq:EmployerBeliefsGenderSensitiveTesting}, as appropriate, to be defined below.

\paragraph{Optimal Hiring.}  In this setting, it is well-known (\textit{e.g.}, \cite{CorbettDaviesEtAl17}) that the employer's optimal rule (from the standpoint of expected payoff maximization) balances benefits (correct choice) against costs (incorrect choice) and is described by a threshold rule of the following form:
\begin{equation}
    \label{Eq:OptimalHiring}
    \delta(a_W, x_W)=
    \begin{cases}
        1 & \text{ if } \Pr[y_W=1\mid a_W, x_W] \; \geq \; \overline{s}(B,\omega), \\
        0 & \text{ otherwise},
    \end{cases}
\end{equation}
where, given the assumptions about the employer's payoffs, the threshold $\overline{s}(B,\omega)$ is defined by the following:
\begin{equation}
\label{Eq:OptimalThreshold}
\overline{s}(B,\omega) \equiv \frac{-\omega}{B-\omega}.    
\end{equation}

\noindent Equation \ref{Eq:OptimalHiring} implicitly characterizes an assessor used by $E$ to calculate the probability that $W$ is qualified, and this assessor is $s= \Pr[y_W=1\mid a_W, x_W]$.  However,   independence of $\overline{s}(B,\omega)$ from the worker's traits ($v_W$) will generally lead optimal hiring by the employer to violate of one or more of the fairness goals discussed above in Section \ref{Sec:Fairness}.  To see this, consider the distribution of traits displayed in Table \ref{Tab:OptimalHiringUnfairExample}, and assume that $\omega=-2$ and $B=+1$. Then, \eqref{Eq:OptimalThreshold} yields the threshold $\overline{s} = 2/3$.

\begin{table}[htbp]
\centering
\begin{tabular}{|c|c|c||c|c|}
\hline
& \multicolumn{2}{c||}{$a_i=M$} & \multicolumn{2}{c|}{$a_i=F$}\\
\hline
$y_i$	& $x_i=0$	& $x_i=1$	& $x_i=0$	& $x_i=1$ \\
\hline
0 		& 0.4			& 0.1     & 0.3			& 0.2			\\
\hline
1 		& 0.1			& 0.4     & 0.2			& 0.3			\\
\hline
\end{tabular}
\caption{An Example of Optimal Hiring Being Unfair\label{Tab:OptimalHiringUnfairExample}}
\end{table}
Given the distribution of traits in Table \ref{Tab:OptimalHiringUnfairExample}, the probability a male with $x_i=1$ is qualified is $\frac{.4}{.4+.1}=\frac{4}{5}$ and the probability a female with $x_i=1$ is qualified is $\frac{.3}{.3+.2}=\frac{3}{5}$.  At hiring threshold $\overline{s}=\frac{2}{3}$ the employer's optimal hiring algorithm will yield  conditional probabilities of the worker being hired given $x_i$ and $a_i$ displayed in Table \ref{Tab:OptimalHiringUnfairExample2}.  While the threshold $\overline{s}$ is independent of $a_i$, the inferences drawn from $x_i$ about $y_i$ are \emph{not} independent of $a_i$.
\begin{table}[htbp]
\centering
\begin{tabular}{|c|c|c||c|c|}
\hline
& \multicolumn{2}{c||}{$a_i=M$} & \multicolumn{2}{c|}{$a_i=F$}\\
\hline
			& $x_i=0$	& $x_i=1$	& $x_i=0$	& $x_i=1$ \\
\hline
$\Pr[\text{hired}\mid x_i,a_i]$	& 0	& 1 & 0 & 0 \\
\hline
\end{tabular}
\caption{An Example of Optimal Hiring Being Unfair, Continued\label{Tab:OptimalHiringUnfairExample2}}
\end{table}
The optimal hiring algorithm in this example will violate both error rate balance and demographic parity.

As discussed in Section \ref{Sec:RationalityAndFairness}, the example in Tables \ref{Tab:OptimalHiringUnfairExample} and \ref{Tab:OptimalHiringUnfairExample2} illustrates the tension between rationality and fairness: rational decision-making by the employer may lead to apparently discriminatory hiring behavior \textit{even if the employer does not care about the worker's sensitive traits}.\footnote{In other words, we have assumed away the possibility of taste-based discrimination.  Allowing for this possibility can be accomplished by allowing $B$ and/or $\omega$ to depend on the worker's permissible and/or sensitive traits (\textit{i.e.}, $B_{v_W}$ and/or $\omega_{v_w}$).}

\subsection{Phelpsian Models of Statistical Discrimination \label{Sec:PhelpsianModels}}

As noted earlier, Phelpsian models of statistical discrimination explicitly allow for differences between the groups, and there are at least two approaches of this form.  The first is based on differences in base rates (\textit{i.e.}, prevalence of qualification) between the groups.  The second is based on the data being more reliable for one group than the other.  As we will show below, these two approaches present different challenges for the attainment of fairness goals.

\subsubsection*{Differences in Prevalences Across Groups} 

To focus squarely on the impact of gender-varying base rates of qualification, we assume for now that $1>p_f>p_m>0$ and that the test result $x$ is a noisy signal of qualification that, conditional on $y$, is distributed as described in Table \ref{Tab:TestResultGenderBlind}.
\begin{table}[ht!]
\centering
\begin{tabular}{|c||c|c|} \hline 
                        &   $y_W=0$         &   $y_W=1$             \\ \hline \hline
$\text{Pr}[x_W=1 \mid y_W]$  &   $\phi$        &   $0$                 \\ \hline
$\text{Pr}[x_W=2 \mid y_W]$  &   $1-\phi$      &   $1-\phi$          \\ \hline
$\text{Pr}[x_W=3 \mid y_W]$  &   $0$             &   $\phi$            \\ \hline
\end{tabular}
\caption{A Gender-Blind Testing Technology \label{Tab:TestResultGenderBlind} }
\end{table}

As noted above, perfect Bayesian equilibrium in this model requires that the employer form correct beliefs about the worker's outcome, $y_W$, conditional on the worker's gender, $a_W$, and test score, $x_W$.  Given the testing technology defined in Table \ref{Tab:TestResultGenderBlind}, the probability that $W$ is qualified conditional on observing $W$'s gender $a_W$ and test score $x_W$ is:
\begin{equation}
\label{Eq:EmployerBeliefsGenderBlindTesting}
\Pr[y_W=1\mid a_W, x_W]=
    \begin{cases}
    0 & \text{ if }    x_W=1, \\
    \frac{p_{a_W}(1-\phi)}{p_{a_W}(1-\phi)+(1-p_{a_W})(1-\phi)}=p_{a_W}  &\text{ if } x_W=2, \\
    1 & \text{ if }x_W=3.
    \end{cases}
\end{equation}

\noindent In words, if $E$ observes a test result of $x=1$ he knows with certainty  that $W$ is unqualified, and if $E$ observes $x=3$ he knows with certainty that $W$ is qualified.  We therefore restrict attention to $E$'s hiring decision in the event that he observes the ``muddled" test result of $x=2$. Letting $d(m)$ and $d(f)$ represent the respective probabilities that a male or a female worker receiving a test score of $x=2$ is hired by $E$, we'll specifically consider decision rules of the following form for any pair of probabilities $(d(m),d(f)) \in [0,1]\times[0, 1]$:\footnote{The optimal rule defined in Equation \eqref{Eq:OptimalHiring}, above, is contained in this family of hiring rules.} 
\begin{equation}
\label{Eq:BanTheBoxHiring}
    \delta(a_W, x_W)=
    \begin{cases}
        1 & \text{ if } x_W=3,\\
        1 & \text{ with probability } d(a_W) \text{ if } x_W=2,\\
        0 & \text{ with probability } 1-d(a_W) \text{ if } x_W=2,\\
        0 & \text{ if } x_W=1.
    \end{cases}
\end{equation} Consequently, $E$ hires those receiving an $x=3$ with certainty and does not hire those receiving an $x=1$.  He hires those receiving an $x=2$ with probability $d(a_W)$.  Before continuing, we can immediately address when the employer's hiring will satisfy anti-classification.  A hiring rule as described by Equation \eqref{Eq:BanTheBoxHiring} satisfies anti-classification if and only if 
    \[
    d(m)=d(f),
    \]  meaning that the probability of being hired conditional on test score is independent of gender.

\paragraph{Fairness in the Phelpsian Model with Different Prevalences.} We are now in a position to summarize under what conditions hiring by the employer will satisfy predictive parity and/or error rate balance.  We'll begin by characterizing the optimal rule, from the employer's perspective.

\begin{itemize}
\item \textit{Optimality.}  By Equations \ref{Eq:OptimalThreshold} and \ref{Eq:EmployerBeliefsGenderBlindTesting}, the optimal decision rule for the employer conditional on observing a score of $x_w=2$ is:
$$d(a_W)=\begin{cases} 1& \text{ if } p_{a_W} \geq \frac{-\omega}{B-\omega},\\
0 & \text{ otherwise.}
\end{cases}
$$ Since the optimal rule is a cutpoint rule, it will satisfy anti-classification whenever the two groups either both do or both don't satisfy the above cutpoint rule.  As we will show below, if anti-classification is met, then, for this gender-blind testing technology, error rate balance is also satisfied.  In the case that the cutpoint is only met for the gender with a higher prevalence, then anti-classification and error rate balance are violated at the optimal rule.  Predictive parity, on the other hand, requires hiring individuals receiving a score of $x_W=2$ probabilistically and at different rates, and is consequently incompatible with a cutpoint rule.  

    \item \textit{Predictive Parity.} If the employer is using a hiring rule of the form described by Equation \eqref{Eq:BanTheBoxHiring}, then employment outcomes cannot satisfy predictive parity.  For example, the positive predictive value of hiring individuals with sensitive trait $a_W$ is equal to the following:
\[
\Pr[y_W=1\mid \delta_W=1, a_W] = \frac{p_{a_W} (\phi + d(a_W) (1-\phi))}{p_{a_W} (\phi + d(a_W) (1-\phi)) + (1-p_{a_W}) d(a_W) (1-\phi)}.
\]
Positive predictive value is increasing in $p_{a_W}$ and decreasing in $d(a_W)$; consequently, equalizing PPV across genders when $p_f>p_m$ requires $\delta(f)>\delta(m)$. In other words, males receiving the test score of $x_W=2$ must be hired at a lower rate than females, and so satisfaction of PPV in this environment necessarily entails a violation of anti-classification unless $d(m)=d(f)=0$, or no false positives are admitted.  Moreover, this relationship between prevalence and hiring rates ($d(f), d(m)$) persists when considering satisfaction of negative predictive value, again requiring that the group with lower prevalence be hired at a lower rate conditional on a score of $x_W=2$.  Consequently, unless certain types of errors are disallowed, predictive parity is incompatible with anti-classification, optimality, and (as we now show) error rate balance.

    \item \textit{Error Rate Balance.}  While predictive parity is incompatible with anti-classification in this setting, the structure of the testing technology implies that any hiring rule that satisfies anti-classification (\textit{i.e.}, $d(m)=d(f)=\tilde{d} \in [0,1]$) necessarily satisfies error rate balance, and vice versa.  This is because, unlike positive and negative predictive values, the true positive and negative rates are independent of group prevalence:
\footnotesize

{\begin{eqnarray*}
TPR= \Pr[\delta_W=1 \mid y_W=1,a_W, \tilde{d}] & = & \frac{\phi+\tilde{d}\cdot (1-\phi)}{\phi+\tilde{d}\cdot (1-\phi)+(1-\phi)(1-\tilde{d})}=\tilde{d}+\phi(1-\tilde{d}),\\
TNR= \Pr[\delta_W=0 \mid y_W=0,a_W,\tilde{d}] & = & \frac{\phi+(1-\phi)(1-\tilde{d})}{\phi+(1-\phi)(1-\tilde{d})+(1-\phi)\tilde{d}}=1-\tilde{d}(1-\phi).
\end{eqnarray*}}

\normalsize
Importantly, the relevant feature of the testing technology that enables satisfaction of error rate balance is the fact that the probability of receiving the ``muddled" test score of $x_W=2$ is independent of both gender and qualification.  We now turn to an environment where this is no longer the case.
\end{itemize}

\subsubsection*{Differences in Information Precision Across Groups\label{Sec:PhelpsianInformationalDiscrimination}} 

In the previous paragraphs we highlighted a setting in which the testing technology was equally informative about male and female outcomes, but discrimination in the form of a violation of predictive parity emerged when prevalence differed between the two genders and the rule was ``gender-blind," satisfying anti-classification.  An observationally similar form of discrimination can occur even if prevalence is identical across the two genders because of gender-based differences in the precision of the testing technology.  Specifically, suppose that $p_m=p_f=\tilde{p} \in [0,1]$, but the quality of the test outcome depends on the worker's sensitive trait, $a_W$, as described in Table \ref{Tab:TestResultGenderSensitive}.
\begin{table}[h!]
\centering
\begin{tabular}{|c||c|c|} \hline 
                            &   $y_W=0$             &   $y_W=1$             \\ \hline \hline
$\text{Pr}[x_W=1 \mid y_W,a_W]$  &   $\phi^{a_W}$      &   $0$                 \\ \hline
$\text{Pr}[x_W=2 \mid y_W,a_W]$  &   $1-\phi^{a_W}$    &   $1-\phi^{a_W}$    \\ \hline
$\text{Pr}[x_W=3 \mid y_W,a_W]$  &   $0$                 &   $\phi^{a_W}$      \\ \hline
\end{tabular}
\caption{A Gender-Sensitive Testing Technology \label{Tab:TestResultGenderSensitive}}
\end{table}  

For this analysis, we assume that $\phi^m\neq \phi^f$. 
Given the testing technology defined in Table \ref{Tab:TestResultGenderSensitive}, the probability that $W$ is qualified conditional on observing $W$'s gender $a_W$ and test score $x_W$ is:
\begin{equation}
\label{Eq:EmployerBeliefsGenderSensitiveTesting}
\Pr[y_W=1\mid a_W, x_W]=
    \begin{cases}
    0 & \text{ if }    x_W=1, \\
    \frac{\tilde{p}\cdot \left(1-\phi^{a_W}\right)}{\tilde{p}\cdot \left(1-\phi^{a_W}\right)+(1-\tilde{p})\left(1-\phi^{a_W}\right)}=\tilde{p}  &\text{ if } x_W=2, \\
    1 & \text{ if }x_W=3.
    \end{cases}
\end{equation}

\paragraph{Fairness in the Phelpsian Model with Different Information Precision.} Again, we are in a position to summarize under what conditions hiring by the employer will satisfy optimality, predictive parity and/or error rate balance.

\begin{itemize}

\item \textit{Optimality.} By Equations \ref{Eq:OptimalThreshold} and \ref{Eq:EmployerBeliefsGenderSensitiveTesting}, the optimal decision rule for the employer conditional on observing a score of $x_W=2$ is:
\[
d(a_W)=\begin{cases} 1& \text{ if } \tilde{p}  \geq \frac{-\omega}{B-\omega},\\
0 & \text{ otherwise.}
\end{cases}
\]
In this case the optimal rule will always satisfy anti-classification.\footnote{Anti-classification is satisfied in this example only because we assume that the probability of receiving a $x=2$ is the same for both qualified and unqualified types.}  However, satisfaction of either error rate balance or predictive parity will require a violation of anti-classification, and error rate balance may not be attainable at all in this setting. We begin with a discussion of predictive parity.

    \item \textit{Predictive Parity.} The positive predictive value of hiring individuals with sensitive trait $a_W$ is equal to the following:
\[
\Pr[y_W=1\mid \delta_W=1, a_W] = \frac{\tilde{p} \cdot (\phi^{a_W} + d(a_W) (1-\phi^{a_W}))}{\tilde{p} \cdot (\phi^{a_W} + d(a_W) (1-\phi^{a_W})) + (1-\tilde{p}) d(a_W) (1-\phi^{a_W})}.
\]It is straightforward to show that both positive predictive value and negative predictive value are increasing in $\phi^{a_W}$, that PPV is decreasing in $d(a_W)$ and that NPV is increasing in $d(a_W)$.  Consequently, predictive parity cannot be satisfied via any hiring strategy ($d(m), d(f)$), and PPV can only be equalized if $d(m)=d(f)=0$. 

    \item \textit{Error Rate Balance.}  Similar to predictive parity, it is straightforward to show that both the true positive rate and the true negative rate are increasing in  $\phi^{a_W}$.  At the same time, the TPR is decreasing in $d(a_W)$ while the TNR is increasing in $d(a_W)$.  Consequently, satisfaction of error rate balance is not possible in this setting.  

\end{itemize}




\paragraph{Fairness \& Rationality in the Phelpsian Model of Statistical Discrimination.} 
The previous sections have shown that the two Phelpsian approaches to modeling group differences (i.e. whether groups differ in their base rates of qualified individuals versus whether the data / testing technology is differentially noisy for the two groups) have real implications for the possibility of achieving fairness goals at all, much less achieving goals that are compatible with optimal decision-making by the employer.  In the former case where groups differ only in prevalence, any decision-rule satisfying anti-classification will necessarily balance error rates for the two groups, and it is possible for optimal-decision making to satisfy anti-classification. In the latter case, it is generically impossible to satisfy either error rate balance or predictive parity at all. 

Setting optimality of the decision rule aside, these points suggest that arguments favoring a ``group-blind" decision rule may be fundamentally misguided if the data describing members of one group are systematically noisier than the data describing another.  And as \cite{ChouldechovaRoth18} note, these environments may be ubiquitous. Consider, for example, predicting college success from SAT scores. Chouldechova and Roth (p.18) write, 
\begin{quote}
``The majority population employs SAT tutors and takes the exam multiple times, reporting only the highest score. The minority population does not. We should naturally expect both that SAT scores are higher amongst the majority population, and that their relationship to college performance is differently calibrated compared to the minority population. But if we train a group-blind classifier to minimize overall error, if it cannot simultaneously fit both populations optimally, it will fit the majority population. This is because---simply by virtue of their numbers---the fit to the majority population is more important to overall error than the fit to the minority population. This leads to a different (and higher) distribution of errors in the minority population."
\end{quote}

\subsection{Arrovian Models of Statistical Discrimination \label{Sec:ArrovianModels}}

Arrovian models of statistical discrimination are similar to Phelpsian models in many ways---the main distinction between the two approaches is that, while Phelpsian models treat the outcomes as exogenously determined, Arrovian models allow for outcomes to be endogenous in the sense that the worker may invest in his or her own outcome, $y_W$.\footnote{\cite{Lundberg83} first extended Phelps's model to also allow for endogeneity of the outcome variable via costly investment in human capital.}  This decision is motivated (at least in part) by a desire to influence the test score, $x_W$, in pursuit of getting hired ($\delta_W=1$).  This endogeneity opens the possibility of \textit{multiple equilibria}.  The existence of multiple equilibria complicates the pursuit of fairness in at least one important way: the hiring rule used by the employer in one equilibrium may satisfy some fairness goals that are violated by the hiring rule used in a different equilibrium.

The existence of multiple equilibria occurs in situations in which SD is ``self-confirming.'' Perfect Bayesian equilibria are based on ``rational expectations'' by the employer, in which he or she has correct beliefs about any given worker's outcome, $y_W$, based on the information available to the employer ($v_W$) at the time he or she makes the hiring decision.  If the employer believes that workers from a given group ($a_W$) are less likely to be qualified ($\Pr[y_W=1\mid a_W]$ is lower), then the employer will require a higher test score to justify hiring individuals from that group.  This can, in some situations, reduce the degree to which investing in effort will increase the probability that a worker from that group will be hired, thereby reducing the probability that a worker from that group will find it in his or her interest to actually invest in becoming qualified.  

\paragraph{\cite{CoateLoury93}.} An excellent analysis of an Arrovian model of statistical discrimination is provided by \cite{CoateLoury93}.  One of several key points of their analysis is that, when the workers' sensitive trait is observed by the employer at the time of making the hiring decision, individuals with different sensitive traits may be treated by the employer differently in the sense that the hiring rule for one group is different from the hiring rule applied to a different group.  This, in turn, leads to each worker's incentive to invest in obtaining qualification endogenously depending on the worker's sensitive trait.  Accordingly, discriminatory behavior by the employer may emerge as a result of the equilibrium played by the employer and worker depending on the worker's sensitive trait (in game theoretic terms, this is referred to as \textit{equilibrium selection}).  For example, it can be the case that the employer believes that women invest in qualification with some positive probability, but that men do not.  In this case, the employer may (correctly) be willing to hire women whose test scores are high enough but (correctly) never hire a male applicant regardless of his or her test score.  \textit{This type of discriminatory equilibrium can emerge even if men and women are otherwise identical.}

\paragraph{The Foundations of Arrovian Models of Statistical Discrimination.}  Arrow's work was building on the Phelps model and, accordingly, Arrovian models of statistical discrimination tend to include the same basic components as Phelpsian models, but also necessarily must include a few more.  For example, in order to incorporate endogeneity of workers' outcomes within an employment setting, one must consider the workers' individual incentives. 
In general, incorporating these incentives leads to the possibility of multiple equilibria.  As mentioned above (Section \ref{Sec:Statistical}), one classic example of equilibrium multiplicity emerges from the possibility of moral hazard: the employer cannot directly observe outcomes, so to the extent that the employer believes that workers with certain characteristics are unlikely to invest in their own outcomes, the employer will be less likely to hire them, \textit{ceteris paribus}, leading in many settings to there being an equilibrium (or multiple equilibria) in which worker with those characteristics do not invest in outcomes and do not get hired.  In general, when multiple equilibria exist, they differ in the hiring rules used by the employer for workers from different groups, so that, \textit{if the employer can observe individuals' sensitive traits, the employer's hiring behavior discriminates between the two groups in some or all of the equilibria.}



\paragraph{Informational Foundations of Arrovian Discrimination.}  As discussed above, Phelpsian models of statistical discrimination can be consistent with discrimination emerging ``simply'' as a result of informational differences between the two groups.  These differences, as portrayed in Section \ref{Sec:PhelpsianInformationalDiscrimination}, emanate from exogenous trait-sensitive differences in the testing technology.  In Arrovian models of statistical discrimination, discrimination is also the result of informational differences.  As opposed to Phelpsian models, the informational differences between groups when discrimination arises in equilibrium in an Arrovian model are \textit{endogenous}.  In a sense, this means that the Arrovian explanation for informationally-induced discrimination is closer to a ``general equilibrium'' explanation than that offered by similar Phelpsian models, because the mechanism of causality is contained within the model.  However, this advantage is mitigated by the lack of a mechanism to explain why the employer and worker (correctly and jointly) believe that their equilibrium behaviors should be conditioned on an exogenous trait that is (by assumption) otherwise \textit{per se} irrelevant to both players.\footnote{Put another way, the equilibrium selection in Arrovian models is effectively playing the same role as the exogenous sensitive trait-sensitivity of the testing technology in the Phelpsian models.}  

\section{Conclusions}

Studies of AF and SD are similarly concerned with ensuring that the decisions affecting individuals are free of discriminatory bias.  However, these fields take very different approaches to the study of fairness. We have argued that the most significant difference between the two approaches is in how they conceive of the outcome-relevant traits of individuals that an algorithm seeks to learn.  Studies of AF tend to conceive of these traits as exogenous to the algorithm itself, with the algorithm simply seeking to identify whether the individuals being classified possess the trait or not.  Models of SD often conceive of these traits as the product of choices made by the individuals being evaluated, and therefore endogenous to the algorithm by which individuals are classified.

In this article we've attempted to draw some connections between these two literatures, demonstrating how and when various well-known notions of AF may be satisfied within a simple model of SD.  We particularly hope that readers will be inspired to pursue the study of how classification algorithms can positively and negatively affect the incentives of individuals to invest in their own defining characteristics.

There are so many directions to pursue in this line of work that move beyond the development of fairness principles that can accommodate how people respond to algorithms.  We are intrigued by the idea of endogenizing the informational environment that an algorithm faces.  In recent work, \cite{PattyPenn21} consider the effect of so-called ``ban the box'' policy proposals which require that employers ignore (or at least delay) considering whether a job applicant has a criminal record when making hiring decisions.  Using a model that blends both the Phelpsian and Arrovian traditions of SD, they show that this reduction of information can (in equilibrium) sometimes benefit not only job applicants, but also the employer.  However, the overall effect is ambiguous: sometimes banning the box can help the employer while \emph{harming} the applicants, and sometimes losing this information can harm \emph{both} the employer and the applicants in equilibrium.   

Another approach to this question would allow individuals to reveal or conceal their own personal information.  If everyone has the opportunity to choose the information available to an algorithm, when will it be (or can it \textit{ever} be) strictly beneficial for a person to conceal information about themselves?  Moreover, there are many environments in which people can manipulate information about themselves.  People can lie, and the more important the information is to an algorithm, the stronger the incentive people will face to manipulate it, and the less informative that information will consequently become. In an example of recent work along these lines, \cite{FrankelKartik22}  demonstrate that decision-makers may have incentives to under-utilize information for reasons that are entirely distinct from those we've previously discussed (e.g. incentivizing skill acquisition, or promoting fair outcomes).  In their model, underutilizing information can actually improve an algorithm's accuracy by reducing peoples' incentives to manipulate data.  It is our hope that this article serves to inspire future research into these, and other, important questions concerning algorithmic design.

\bibliography{john-fairness}

\end{document}